\documentclass[aps,  12pt]{revtex4}
\usepackage{graphicx,subfigure}
\usepackage{hyperref}
\usepackage{epsfig}
\usepackage{xcolor}
\usepackage{amsmath}
\usepackage{amssymb}
\usepackage{amsthm}
\usepackage{bbold}
\begin{document}
	\title{Negative refractive index, Perfect Lens and Ces\`{a}ro convergence }
	\author{Yuganand Nellambakam and  K. V. S. Shiv Chaitanya }
	\email[]{ chaitanya@hyderabad.bits-pilani.ac.in}
	\affiliation{Department of Physics, BITS Pilani, Hyderabad Campus, Jawahar Nagar, Shamirpet Mandal,
		Hyderabad, India 500 078.}
	\begin{abstract}
		In this letter, we show that the restoration of evanescent wave in perfect lens obeys a new kind of convergence known as Cesaro convergence. Cesaro convergence
		allows us to extend the domain of convergence that is analytically continuing to the complex plane in terms of Riemann zeta function. Therefore, from the properties of Riemann zeta function we show that it is not possible to restore the evanescent wave
		for all the values of $r_z'$, [here $r_z'$ is complex]. The special value, that is, $r_z'$ = 1=2+ib refers to the non-existence of evanescent wave, is the physicists proof of Riemann Hypothesis.
		
	\end{abstract}
	
	\maketitle
	
J. B. Pendry has shown that when the refractive index '$n$' of a lens is negative, the evanescent waves can be restored in such a way that they play a role in image formation \cite{len}. The processes of restoration of  evanescent waves gives rise to a batter resolution of the image, with the optical resolution less than
\begin{equation}
\Delta\approx\frac{2\pi}{k_{max}}=\frac{2 \pi c}{\omega}=\lambda.
\end{equation}

The results of perfect lens were challenged by t'Hooft \cite{hoo} and  Williams’s \cite{will}.  Following claims were questioned in that reference \cite{hoo} on causality argument, amplitude transmission coefficient of the slab and summation over various terms in the geometrical series that do not converge.  In another reference \cite{will} following claims were criticized, that is,  perfect lens can focus light onto an area smaller than a square wavelength, complexity of  refractive index, and  the evanescent field is not  distinguished between it and the field of a plane wave.

The reply to both criticisms were given in  references \cite{penr1} and \cite{penr} respectively. In this letter, we would like to address the issue of  geometric convergence, in the wake of recent work by the authors \cite{kvs}, that is, metamaterials obey new kind of convergence known as Ces\`{a}ro convergence.  In the  reply to  geometric convergence \cite{penr1}it is stated that: "\textit{ summation of an infinite series is a standard technique in multiple scattering theory, provided the summation is made exactly, the correctness of answer survives the formal divergence of the series, and the result that can be understood through arguments of analytic continuation. The topic is dealt with in  in Ref. \cite{pendr} and more
completely in \cite{pendr1}.}" It should be noted that the author in reply has rightly pointed out that the divergence of geometric series is well understood through analytic
continuation, but the author has not given how the analytic
continuation is achieved.  
Therefore, this is to rectify the inconsistency in argument by showing that geometric series does converge through a special kind of convergence known as the  Ces\`{a}ro convergence for values out side the range of geometric convergence. This Ces\`{a}ro convergence allows us to analytic continue the series into complex plane and the summation is given in terms of Riemann zeta function. This, in turn, makes the refractive index complex and answers  the second question raised in the reference \cite{will}.

To demonstrate that the perfect lens obeys the Ces\`{a}ro convergence we follow the procedure and definitions given in the reference \cite{len}.	It is well known, in classical optics, that the image is formed in a 2D plane, also known as focal plane as the object evolves along the third dimension, perpendicular to the focal plane, into an image. The electric field in focal plane  described by the 2D Fourier components is given by
	\begin{equation}
	E(r,t) = \sum_{\sigma,k_x,k_y}\textbf{E}_{\sigma}  \left( k_x,k_y \right)exp \left( ik_zz+ik_xx+ik_yy-i\omega t \right)
	\end{equation}
	and the wave vector along z-axis is defined for $\omega^2 c^{-2}>k_x^2+k_y^2,$  as
	\begin{equation}
	k_z = +\sqrt{\omega^2 c^{-2}-k_x^2 - k_y^2}, \;\;\;\ 
	\end{equation} These 2D Fourier components are used for  image formation. It is also well known that when $ \omega^2 c^{-2}<k_x^2+k_y^2$  the wave vector 
	\begin{equation}
	k_z = +i\sqrt{k_x^2 + k_y^2-\omega^2 c^{-2}}, \;\;\;\ 
	\end{equation}
gives rise to the evanescent waves. Readers should note that the 2D Fourier components are the phenomenona of far field, and the evanescent waves are the near field phenomena. Therefore, the evanescent wave decay much before the image formation when the refractive index is positive. The evanescent wave is restored if the refractive index is negative.
The transmission  coefficient is given by 
\begin{equation}
T=tt^{\prime}=exp\Bigg(ik^{\prime}_z d\Bigg)=exp\Bigg(-i\sqrt{\omega^2 c^{-2}-k_x^2 - k_y^2}\Bigg)
\end{equation}
with 
\begin{equation}
k^{\prime}_z = -\sqrt{\omega^2 c^{-2}-k_x^2 - k_y^2}
\end{equation}
by matching the fields at boundary
\begin{equation}\label{tp}
t=\frac{2\mu k_z}{\mu k_z+k^{\prime}_z},\;\;\;\;\;\;r=\frac{\mu k_z - k^{\prime}_z }{\mu k_z + k^{\prime}_z}
\end{equation}
and the waves inside  medium give
\begin{equation}\label{tpp}
t^{\prime}=\frac{2 k^{\prime}_z}{k^{\prime}_z+\mu k_z},\;\;\;\;\;\;r^{\prime}=\frac{k^{\prime}_z - \mu k_z }{k^{\prime}_z + \mu k_z}
\end{equation}
the transmission through both surfaces of the slab computed by summing the multiple scattering events
\begin{eqnarray}\label{th}
T_S &=& tt^{\prime}exp\Bigg(ik^{\prime}_zd\Bigg)
+tt^{\prime}r^{\prime2}exp\Bigg(3ik^{\prime}_zd\Bigg)\nonumber\\
&&+tt^{\prime}r^{\prime4}exp\Bigg(5ik^{\prime}_zd\Bigg)+\cdots \nonumber\\
&=&\frac{tt^{\prime}exp\Bigg(ik^{\prime}_zd\Bigg)}{1-r^{\prime2}exp\Bigg(2ik^{\prime}_zd\Bigg)}
\end{eqnarray}
by taking the limit $\mu=\epsilon=-1$ and by substituting the transmission and reflection coefficients the following result will be obtained 
\begin{eqnarray}
\lim_{\substack{\varepsilon\rightarrow-1 \\ \mu \rightarrow-1}}T_S&&= \lim_{\substack{\varepsilon\rightarrow-1 \\ \mu \rightarrow-1}}\frac{tt^{\prime}exp\Bigg(ik^{\prime}_zd\Bigg)}{1-r^{\prime2}\;exp\Bigg(2ik^{\prime}_zd\Bigg)} \nonumber\\
&&= exp\Bigg(-ik^{\prime}_zd\Bigg)=exp\Bigg(-ik_zd\Bigg)
\end{eqnarray}
Similar results are obtained for reflection coefficient. 

The point of contest is the series in equation (\ref{th}) does not converge for all values of the $r'^2$ which restricts $r'^2<1$. It can be seen explicitly by taking $k'_zd=\pi$ which reduces the series to be 
\begin{eqnarray}\label{th1}
T_S && = -tt^{\prime}(1+r^{\prime2}+r^{\prime4}+\cdots )\nonumber\\
&&=\frac{tt^{\prime}}{1-r^{\prime2}}
\end{eqnarray}
As pointed out in reference \cite{pendr1} the value of $\epsilon$ and $\mu$ are complex, hence we consider $r'=i$. We justify this value of  $r'$ as we are going show that the geometric series outside the range of convergence obeys Ces\`{a}ro convergence which in turn makes $r'$  complex. The value of $r'=i$ implies $\mu=\frac{1-i}{1+i}=(1-i)^2/2$ with $k_z'=k_z$ in equation (\ref{th})
then the series in equation (\ref{th1}) reduces to  
\begin{eqnarray}\label{th2}
T_S && = -tt^{\prime}(1-1+1-1+1\cdots )
\end{eqnarray}
In literature, the series in equation (\ref{th2}) is known as Grandi's series. Grandi's series  does not obey the regular  geometric convergence that is the sum to infinity $\sum_{0}^{\infty}x^n=\frac{1}{1-x}$  is not defined.  It is well known that for a geometric series to converge, the value of  $x$ should lie in the range $-1<x<1$; here $-1$ and $1$  are also excluded. In Grandi's series, the value of $x$ is $x=-1$ as given
\begin{equation}\label{cs1}
Q=1-1+1-1+1.......=\sum_{n=0}^{+\infty}Q_j=\sum_{n=0}^{+\infty}(-1)^n.
\end{equation}
It is interesting to note that Ramanujan \cite{ram} has used the value of $x=-1$ in the geometric series sum to infinity $\sum_{0}^{\infty}x^n=\frac{1}{1-x}$ and obtained the value of $\sum_{0}^{\infty}x^n=1/2$. The value of $1/2 $ as the sum to infinite series in equation (\ref{cs1}) is justified if we assume that the Grandi's series obeys Ces\`{a}ro convergence. We give a brief description of Ces\`{a}ro convergence bellow: For a geometric series to converge the sequence  of partial sums should converge to real number.  The  sequence of partial sums for Grandi's series gives
\begin{eqnarray}\label{prg}
P_0&=&Q_0=1,;\;\;\;\ P_1=Q_0+Q_1=0,;\;\;\;\ \\ P_2&=&Q_0+Q_1+Q_2=1, ;\;\;\;\ \\ P_3&=&Q_0+Q_1+Q_2+Q_3=0,......
\end{eqnarray}
It is clear for equation (\ref{prg}) that the  sequence of partial sums  does not converge to a real number. But, the sum to infinity of geometric series gives $\sum_{0}^{\infty}(-1)^n=\frac{1}{1-(-1)}=\frac{1}{2}$, a real number. The RHS converges and LHS diverges, hence for consistency we consider the averages of partial sums, that is,
\begin{eqnarray}\label{prg1} 
\frac{P_0}{1}=1,;\;\;\;\ \frac{P_0+P_1}{2}&=&\frac{1}{2},\\
\frac{P_0+P_1+P_2}{3}&=&\frac{2}{3}, \\ \frac{P_0+P_1+P_2+P_3}{4}&=&\frac{2}{4}
\end{eqnarray} and so on. Sequence of the average of partial sums gives \begin{equation}\label{gh}
P_n = \frac{1}{n}\sum_{k=1}^{n}Q_n= \frac{1}{1},\frac{1}{2},\frac{2}{3},\frac{2}{4},\frac{3}{5},\frac{3}{6},\frac{4}{7},\frac{4}{8},\frac{5}{9},\frac{5}{10},...
\end{equation}
The equation (\ref{gh}) is recasted as
\begin{equation}
P_n=\begin{cases}
\frac{1}{2}, &for \;\;n \;\;odd\\
\frac{1}{2}+\frac{1}{2n+2}, &for\;\; n \;\;even
\end{cases}
\end{equation}
which, as n goes to infinity, converges to $\frac{1}{2}$. From the equation (\ref{gh}), it is clear that the average of partial sums converges to a real number, and this kind of convergence is known as Ces\`{a}ro Convergence.
A series $\sum_{j=0}^{n} a_j $ is  Ces\`aro summable if this satisfies the following theorem:

\textbf{Theorem 1} Suppose  $\sum_{j=0}^{n} a_j $ is a convergent series with sum, say L. Then $\sum_{j=0}^{n} a_j $ is Ces\`aro summable to L. 
\begin{equation}\label{def1}
\lim_{n\rightarrow\infty}s_n = L \in \mathbb{R}\;\;\;\;\Rightarrow\;\;\;\lim_{n\rightarrow\infty}\sigma_{n} = L \in \mathbb{R}.
\end{equation}
The proof is given in \cite{cso}. Following are the properties of  Ces\`aro sums:
If $\sum_n a_n$ = A and $\sum_n b_n$ = B are convergent series, then \\
i. Sum-Difference Rule:  $\sum_n(a_n\pm b_n)$ = $\sum_n a_n \pm \sum _n b_n$ = A $\pm$ B \\
ii. Constant Multiple Rule : $\sum_n c\; a_n$ = c $\sum_n a_n$ = cA for any real number c.\\
iii. The product of $AB=\sum_n a_n\sum_n b_n$  also as Ces\`aro sums. 

The sequence in equation (\ref{cs1}) has two possibilities; one case is that the sequence ends with an even number of terms, and the other argument is that series ends with an odd number of terms.  For even of terms, the partial sums add to $s_{2n} = 0$, and for the odd number of terms, the partial sums add to $s_{2n+1} = 1$; then the average of even and odd is $1/2$. Then, by followings the  theorem1 we get
\begin{eqnarray}
\sigma_{2n+1} &=& \frac{1}{2n+1}(1+0+1+0+...+1)=\frac{n+1}{2n+1} \label{odd}\\
\sigma_{2n} &=& \frac{1}{2n}(1+0+1+0+...+0) =\frac{1}{2}\label{even}
\end{eqnarray}
Then, by applying theorem 1 we get 
\begin{equation}\label{jk}
\lim_{n\rightarrow\infty}\; \sigma_{2n} = \lim_{n\rightarrow\infty}\sigma_{2n+1} = \frac{1}{2}.
\end{equation}
Hence, substituting (\ref{jk}) in equation (\ref{th2}), also we get the values of $tt'=2$ using equations (\ref{tp}) and (\ref{tpp}) with $\mu=(1-i)^2/2$ and $k_z=k_z'$ gives
\begin{eqnarray}\label{thp2}
T_S && = -1
\end{eqnarray}
Therefore, we recover the result of in the reference \cite{len} for $k_z'd=\pi$ and the transmission coefficient in equation (\ref{th}) obeys Ces\`{a}ro Convergence. If a series is Ces\`aro summable, the argument in the series  becomes complex \cite{sas}. Thus, Ces\`{a}ro Convergence allows us to analytically continue the geometric series to the complex plane. A similar results can be obtained for reflection coefficient as shown in the reference \cite{len}. It should be noted, we have only shown that Ces\`{a}ro Convergence is satisfied for a particular value in equation (\ref{th}). To derive a more general result we analytically continue the series in equation (\ref{th}) by  Riemann Zeta($\zeta$) function. We obtain this result as Ces\`{a}ro Convergence is intimately connected to the Riemann Zeta($\zeta$) function through the  Dirichlet eta function
\begin{eqnarray}
\eta(s) &=& 1-\frac{1}{2^s}+\frac{1}{3^s}-\frac{1}{4^s}+\frac{1}{5^s}+....\\&=&\left(1+\frac{1}{2^s}+\frac{1}{3^s}+\frac{1}{4^s}+\frac{1}{5^s}+..\right)
\nonumber \\&&-\left(\frac{2}{2^s}+\frac{2}{4^s}+\frac{2}{6^s}+\frac{2}{8^s}+..\right)\nonumber\\
&=& \zeta(s)-\frac{1}{2^{s-1}}\zeta(s)= (1-2^{1-s})\zeta(s).
\end{eqnarray}  where Dirichlet eta function is given by
\begin{equation}\label{dr}
\eta(s)=\sum_{k=1}^{+\infty}\frac{(-1)^k}{k^s}=1-\frac{1}{2^s}+\frac{1}{3^s}-\frac{1}{4^s}+\frac{1}{5^s}+......
\end{equation}
The Grandi's series is obtained from Dirichlet eta function for the value $s=0$.
To obtain  Riemann Zeta($\zeta$) function for the transmission coefficient we map the value of $r'^2$ to $e^{-2r}$ here $n=1,2,\cdots$ this reduces the equation (\ref{th}) to be
\begin{eqnarray}\label{tha}
T_S &=& tt^{\prime}(e^{iK_zd}+e^{-2r}e^{i3K_zd}
+e^{-4r}e^{i5K_zd}+\cdots)\nonumber\\
&=&\frac{tt^{\prime}exp\Bigg(ik^{\prime}_zd\Bigg)}{1-e^{2r}exp\Bigg(2ik^{\prime}_zd\Bigg)}
\end{eqnarray} 
The series in equation (\ref{th}) is written as
\begin{eqnarray}\label{tha1}
T_S &=& tt^{\prime}\sum_{s=0}^nexp(-2sr+(2s+1)ik'_zd)
\end{eqnarray} 
Euler derived Riemann Zeta($\zeta$) function through the following series
\begin{equation}\label{eu}
e^{-y}-e^{-2y}+e^{-3y}-e^{-4y}+\cdots=\frac{1}{e^y+1},
\end{equation}
which converges for all y $>$ 0. For $y=0$ the equation (\ref{eu})  reduces to Grandi's series. 

 To obtain Riemann Zeta function  for the transmission coefficient  $r$ in   equation (\ref{tha1}) be complex and should reduced to equation (\ref{eu}). As $r$ in   equation (\ref{tha1}) has to be complex we assume $r=q+i\pi/2$ this gives
\begin{eqnarray}\label{tha4}
T_S &=& tt^{\prime}\sum_{s=0}^nexp(-2s(q+i\pi/2)+(2s+1)ik'_zd))
\end{eqnarray} 
this reduces to 
\begin{eqnarray}\label{tha41}
T_S &=& tt^{\prime}e^{ik'_zd}\sum_{s=0}^n(-1)^sexp(-2s(q+ik'_zd))
\end{eqnarray} 
Taking $y=(q+ik'_zd)/2$ we obtain equation (\ref{eu}).

By differentiating equation (\ref{eu}) $n$ times, we obtain
\begin{equation}
1^ne^{-y}-2^{n}e^{-2y}+3^{n}e^{-3y}-4^ne^{-4y}+\cdots=(-1)^n\frac{d^n}{dy^n}\left(\frac{1}{e^y+1}\right),
\end{equation}
which again converges for any y $>$ 0. Then by expanding the function  $1/(e^y+1)$ around $y=0$ using
Taylor series, we get
\begin{equation}\label{fh}
\frac{1}{e^y+1} = \sum_{k=0}^{\infty} a_k y^{k}
\end{equation}
Euler, by assuming $k$ to be complex in equation (\ref{fh}) derived the following functional form of  Riemann Zeta function \cite{har,divr}
\begin{equation}\label{fh1}
\zeta(1-s)=2(2\pi)^{-s}cos\left(\frac{s\pi}{2}\right)\Gamma(s)\zeta(s),
\end{equation}
where $s$ is complex. Readers should note that  $s$ is related to $k$ in the  equation (\ref{fh}). For more details refer to \cite{har,divr}. Thus, the transmission coefficient in terms of Riemann zeta function is given by
\begin{eqnarray}\label{tha41}
T_S &=& 2(2\pi)^{-s} tt^{\prime}cos\left(\frac{s\pi}{2}\right)\Gamma(s)\zeta(s) e^{ik'_zd}.
\end{eqnarray} 

 Therefore, we have shown that the transmission coefficient in equation (\ref{th}) obeys a new kind of convergence known as Ces\`{a}ro   convergence and this, in turn, has allowed us to analytically continue the series in terms of Riemann Zeta function. Similarly, another series given in the reference \cite{len} also can be written in terms of  Riemann Zeta function.
 
By applying zeros of  Riemann Zeta function we find some interesting behavior of the perfect lens.  Riemann Zeta function has trivial zeros at $\zeta(-2s)$ and non-trivial zeros, which are conjectured Riemann known as Riemann hypothesis on the half line, that is, $\zeta(1/2+iy)$.  by   question would arise when $\zeta(s)=0$. Whenever the Riemann Zeta function is zero, we find that the transmission coefficient in equation (\ref{th}) is zero that is evanescent wave vanishes. Thus, evanescent wave vanishes at the trivial and non-trivial zeros of the Riemann Zeta function. Therefore, we have shown that by analytically continuing the transmission coefficient to the complex plane evanescent wave will not be restored for all values of $r'$ in equation (\ref{th}). More interesting case will be the non-trivial zeros of Riemann Zeta function. Absence of evanescent wave on the half line is a restatement of Riemann hypothesis. Therefore, if one experimentally demonstrates the absence of  evanescent wave on the half line, it  is the proof of physicist of Riemann hypothesis.     
 
In conclusion, we show that the restoration of  evanescent wave in perfect lens obeys a new kind of convergence known as Ces\`{a}ro  convergence. This Ces\`{a}ro  convergence allow us to extend the domain of convergence that is analytically continuing, to the complex plane in terms of Riemann zeta function. By analytical continuation of transmission coefficient  to complex plane we have successfully answered the question raised on the convergence of geometric series in reference \cite{hoo}. This, in turn, also answers the question on complexity of refractive index in the reference \cite{will}.  Thus, we have shown from the properties of Riemann zeta function that it is not possible to restore the evanescent wave for all values of $r'$, [here $r'$ is complex]. The special value, that is, $r'=1/2+ib$ establishes the non-existence of  evanescent wave, which is the physicists proof of Riemann Hypothesis.  
\section*{Acknowledgments}
KVSSC acknowledges the Department of Science and Technology, Govt of India (fast-track scheme (D. O. No: MTR/2018/001046)), for financial support.


\begin{thebibliography}{99}
	\bibitem{len} J. B. Pendry,  \href{https://doi.org/10.1103/PhysRevLett.85.3966}{\textcolor{blue}{Phys.Rev.Lett.}} \textbf{85}, 3966 (2000).
	\bibitem{hoo}G.W. ’t Hooft, Phys. Rev. Lett. 87,
	249701 (2001).
	\bibitem{will}J. M. Williams,  Phys. Rev. Lett. 87,
	249703 (2001).	
	\bibitem{penr1}John Pendry
	Phys. Rev. Lett. 87, 249702 (2001)
	\bibitem{penr}John Pendry
	Phys. Rev. Lett. 87, 249704 (2001)
	\bibitem{pendr}J. B. Pendry, Low Energy Electron Diffraction (Academic
	Press, London, 1974).
	\bibitem{pendr1} R. G. Newton, Scattering Theory of Waves and Particles
	(McGraw-Hill, New York, 1966).
	\bibitem{kvs}Yuganand Nellambakam and  K. V. S. Shiv Chaitanya, AIP Advances 10, 045127 (2020); https://doi.org/10.1063/1.5144629.
		\bibitem{ram} Ramanujan's Notebooks, retrieved January 26, (2014)
	\bibitem{cso}E. Ces\`{a}ro, Bull. Sci. Math., 14 : 1 (1890) pp. 114–120
	\bibitem{sas}Stephen Semmes, “Sums and averages,” \href{https://arxiv.org/pdf/1008.2467v1.pdf}{\textcolor{blue}{arXiv.org}}  (2010)
	\bibitem{har}Hardy, G.H. Divergent Series. Clarendon Press, Oxford. 1949.
	
	\bibitem{divr}Bryden Cais, \href{https://www.math.arizona.edu/~cais/Papers/Expos/div.pdf}{\textcolor{blue}{lecture notes}}
	
	
\end{thebibliography}
\end{document}